\documentclass[aps,prd,onecolumn,notitlepage]{revtex4-1}
\usepackage{graphics,psfrag}
\usepackage{graphicx,psfrag}
\usepackage{epic,eepic}
\usepackage{color} 
\usepackage{epsfig}
\usepackage{latexsym}
\usepackage{pstricks}
\newcommand{\be}{\begin{equation}}
\newcommand{\ee}{\end{equation}}
\newcommand{\bear}{\begin{eqnarray}}
\newcommand{\eear}{\end{eqnarray}}
\begin{document}

\title{Flavor dependence of normalization constant for an infrared renormalon}
\author{Taekoon Lee}
\email{tlee@kunsan.ac.kr}
\affiliation{Department of Physics, Kunsan National University, 
Kunsan 573-701, Korea}


\begin{abstract}
An ansatz is proposed for the flavor dependence of the normalization constant for the
 first IR renormalon in heavy quark pole mass.
\end{abstract}

 

\maketitle

The ultraviolet (UV) and infrared (IR) renormalons in field theory are fascinating objects 
that give rise to factorially growing large order behavior in perturbation theory.
In quantum chromodynamics (QCD) the IR renormalons can be understood within the framework 
of operator product expansion (OPE) \cite{parisi0,parisi1,david1,david2,nsvz1,nsvz2}. 
An IR renormalon in a Wilson coefficient in the OPE 
causes ambiguity in the resummed asymptotic series of the Wilson coefficient, which is  
to be cancelled by the vacuum condensate of an higher dimensional operator. 
This cancellation of the ambiguities in the resummation and vacuum condensate 
determines the nature of the renormalon singularities in the Borel plane. 
Specifically, the functional form of the condensate in the strong coupling $\alpha_s$, 
which can be determined up to an overall constant by renormalization group (RG) equation, 
determines the power of singularity as well as its location \cite{mueller}. 
However, the residue of the singularity, 
which is the normalization constant of the renormalon-caused large order behavior, 
is not known, but it can be calculated by a perturbation method \cite{lee1,lee2}.
 While the normalization
 can be expressed in a convergent series, with finite order perturbation it can only be 
 calculated approximately. Its exact form is a nonperturbative quantity and so far
  there is no known way to calculate it. It would thus be very interesting 
  if we somehow find the exact form for the normalization.

Our purpose in this Letter is to present an ansatz for the normalization for the first 
IR renormalon in the heavy quark pole mass. In this case the perturbation method for the 
normalization yields a series that converges rather quickly, which allows to determine 
the normalization within a few percent of uncertainty using the known first three 
perturbative coefficients \cite{pineda,lee3}. 
The reason for this rapid convergence and accuracy may 
lie with following two facts. First, the renormalon singularity is relatively soft, 
with its singularity of $(1-2b)^{-(1+\beta_1/2\beta_0^2)}$, for example, compared to 
that of the first IR renormalon in Adler function, which is 
of $(1-b/2)^{-(1+2\beta_1/\beta_0^2)}$, where $\beta_0,\beta_1$ are
the first two coefficients of the beta function and $b$ is the complex variable for
 the Borel plane. With vanishing flavor number $N_f=0$, for example, 
 the singularities are $(1-2b)^{-1.42}$ and $(1-b/2)^{-2.69}$, respectively.  
 Secondly, the IR renormalon  is the closest singularity to the origin in the Borel plane. 
 This condition is important because the normalization in the perturbation method 
 is evaluated on the boundary of convergence disk and for that to work the
  singularity should be closer to the origin than any other singularities. 
  Of course, any renormalon singularity can be moved by a conformal 
  mapping to be the closest one to the origin, this step is not required 
  with the first IR renormalon for the pole mass,  because it is already 
  the closest one, and this seems to help the convergence. 
  The normalization constant obtained from perturbation method was 
  confirmed by recent lattice calculation of large order behaviour 
  of static energy \cite{bali0,bali1}.

To get the ansatz let us assume that the condensate in the OPE is of dimension $n$ and is 
proportional to
\bear
\Lambda_{\rm QCD}^n\sim 
e^{-\frac{n}{2\beta_0\alpha_s}}\alpha^{-\frac{n\beta_1}{2\beta_0^2}}(1+{\cal O}(\alpha_s))\
\label{condensate}
\eear
and the associated renormalon singularity is of the form:
\bear
\frac{\cal N}{(1-b/b_0)^{1+\nu}}\,.
\eear
Then the imaginary part, which is ambiguous, of the Borel integral
\bear
{\rm Im}\left[\frac{1}{\beta_0}\int_0^\infty e^{-b/\beta_0\alpha_s}\frac{\cal N}{(1-b/b_0)^{1+\nu}}db\right]
=\pm {\cal N} \sin(\nu\pi)\Gamma(-\nu) (b_0/\beta_0)^{1+\nu}
 e^{-b_0/\beta_0\alpha_s} \alpha_s^{-\nu}(1+{\cal O}(\alpha_s))
\label{ambi}
\eear
 is to be cancelled by ambiguity of the form (\ref{condensate}). 
 Thus identifying (\ref{ambi}) with (\ref{condensate}) we get
 \bear b_0=n/2\,,\,\,\, \nu=n\beta_1/2\beta_0^2\,,\,\,\,
 {\cal N}=\frac{C (\frac{2\beta_0}{n})^{\nu}}{\sin(\nu\pi)\Gamma(-\nu)}\,,
 \label{rel}
 \eear
 where $C$ is an unknown proportionality constant. Now putting 
 \bear
 C=C_0f(\nu)
 \label{cnu}
 \eear
  with $f(0)=1$,
 where $C_0$ is determined so that the normalization agrees with 
 that from the large-$\beta_0$ approximation at $\nu=0$,
 we get
 \bear
 {\cal N}=\frac{C_{0}f(\nu) (\frac{2\beta_0}{n})^{\nu}}{\sin(\nu\pi)\Gamma(-\nu)}\,.
 \eear
 
 We shall now consider the normalization for the pole mass, for which $n=1$. 
 While in this case the ambiguity in Borel summation is not canceled by a condensate 
 of dimension one operator but by the static inter-quark potential \cite{beneke4}, 
 the ambiguity is nevertheless of the form (\ref{condensate}) 
 and the relations  (\ref{rel}) are still valid.
 For this case
 the large-$\beta_0$ approximation 
 fixes $C_0$  in the $\overline{\rm MS}$ scheme as \cite{braun,bigi,beneke1}
 \bear 
 C_0=-C_F e^{5/6}\,,
\eear 
where $C_F=4/3$ for QCD.
Of course $f(\nu)$ is not known, and we can only guess it by comparing
 the normalization with values from the perturbation method. This gives
\bear
f(\nu)=4^{-\nu}
\label{fnu}
\eear
and, finally,
\bear
 {\cal N}=-\frac{C_F e^{\frac{5}{6}}(\frac{\beta_0}{2})^{\nu}}{\sin(\nu\pi)\Gamma(-\nu)}
 \,,
 \label{ansatz}
 \eear  
   for the $\overline{\rm MS}$ scheme, where $\nu=\beta_1/2\beta_0^2$. 
   This is our proposed ansatz.
  
  \begin{table}[t]
\begin{tabular}[t]{lcccccc}\hline\hline
$N_f$&0&1&2&3&4&5 \\ \hline
${\cal N}_{\rm pert.}$&0.6081&0.5981&0.5867&0.5730&0.5551&0.5288\\ \hline
${\cal N}/{\cal N}_{\rm pert.}$&1.0047&0.9979&0.9973&1.0081
&1.0403&1.116\\ \hline\hline
\end{tabular}
\caption{Comparison of the ansatz (\ref{ansatz})
with normalization from perturbation method for varying number of flavors.}
\label{table1}
\end{table}

How well does this ansatz work? Table \ref{table1} compares 
the ansatz with the normalization 
from the perturbative method for varying number of flavors, 
the details of which will be discussed shortly. 
The agreement is quite impressive for up to $N_f=4$. 
It is to be noted that the perturbative method tend to work better for smaller $N_{f}$.
 Of course, this does not mean that the ansatz
 is necessarily correct. One test may be to expand (\ref{ansatz}) for small $\nu$ and
compare it with subleading corrections in large-$\beta_0$ 
approximation, for which only partial result exists \cite{bene}. 
Expanding (\ref{ansatz}) at $\nu=0$, we have
\bear
{\cal N}= \frac{C_F e^{5/6}}{\pi}[1+(\gamma_E +\log(\beta_0/2))\nu+{\cal O}(\nu^2)]\,.
\label{exp}
\eear
 The Euler constant term agrees with the corresponding term in \cite{bene}. 
 It is worth noting that the two 
 transcendental numbers $\pi$ and $\gamma_E$ in (\ref{exp}) both arise from 
\bear
\sin(\nu\pi)\Gamma(-\nu){}
\label{singamma}
\eear
in the denominator in (\ref{ansatz}). This is a strong indication 
that the term should be, at least, part of
the exact form for the normalization.

It may be tempting to use the same idea to make an ansatz for 
other renormalons, for example, such as the
first IR renormalon in the Adler function, which is associated with 
the gluon condensate, for which $n=4$. However, there is no
 accurate estimate of the normalization for the Adler function; 
 The convergence from the perturbative method is not fast \cite{lee-largeorder}.
 Without accurate numerical estimates for the normalization it is impossible to infer 
 $f(\nu)$ for the Adler function and that blocks such an attempt. Nevertheless,
  if we assume (\ref{fnu}) is applicable to the Adler function as well, then we 
  have an ansatz  for the first IR renormalon of the Adler function
 \bear
 {\cal N}=-\frac{\frac{3}{4}C_F e^{\frac{10}{3}}(\frac{\beta_0}{8})^{\nu}}{\sin(\nu\pi)\Gamma(-\nu)}
 \,,
 \label{ansatz-Adler}
 \eear  
where $\nu=2\beta_1/\beta_0^2$, and the following large-$\beta_0$ result is used \cite{beneke1,beneke3,broadhurst}:
\bear
C_0=\frac{3}{4}C_F e^{\frac{10}{3}}\,.
\eear

It is interesting to see how the ansatz compares with perturbative results, 
even though the convergence for them is not that good and uncertainty is too 
large for accurate comparison. From \cite{lee-largeorder}, which uses five-loop Adler function,
 the normalization from perturbation
is given as $0.287,0.251,0.208,0.154$ for $N_{f}=0,1,2,3$, respectively, 
and the corresponding numbers
from the ansatz are $0.327,0.300,0.282,0.277$.
As expected the ratios of corresponding numbers are not close to unity, 
but what is remarkable is that 
numbers from the ansatz and perturbation fall in the same ballpark. 
This is not so a trivial point because  there is an order of magnitude difference
between the large-$\beta_0$ result
\bear
\frac{e^{\frac{10}{3}}}{\pi}=8.9
\eear
and the numbers from the perturbative method, and there is no obvious 
reason that a function of such  complex form as (\ref{ansatz-Adler}) 
should give values that are of same magnitude as the perturbative numbers.

We now show how ${\cal N}_{\rm pert.}$ in Table \ref{table1} were obtained.
The bilocal expansion, which interpolates
 the two expansions about the origin and the renormalon singularity, of 
the Borel transform $\tilde m(b)$ for the pole mass is given in the form \cite{lee3,lee4}:
\bear
\tilde m(b)=\sum_{n=0}\frac{h_{n}}{n!}\left(\frac{b}
{\beta_{0}}\right)^{n}+\frac{\cal N}{(1-2b)^{1+\nu}}(1+\sum_{i=1} c_{i}(1-2b)^{i})\,,
\label{bilocal}
\eear
with which the Borel summed pole mass $m_{{\rm BR}} $ is given by
\bear
m_{\rm BR} =m_{\overline{\rm MS}}\left[1+{\rm Re}\left(\frac{1}{\beta_{0}}  \int_{0}^{\infty} 
e^{-b/\beta_{0}\overline{\alpha}_{s}} \tilde m(b) db\right)\right]\,,
\eear
where $\overline{\alpha}_{s}\equiv\alpha_{s}(m_{\overline{\rm MS}})$\,.
The perturbative form of the Borel transform is
\bear
\tilde m(b)= \sum_{n=0}\frac{p_{n}}{n!}\left(\frac{b}
{\beta_{0}}\right)^{n}\,,{}
\label{pertBorel}
\eear
which gives the perturbative expansion of the pole mass:
\bear
m_{\rm pole}=m_{\overline{\rm MS}}(1+\sum_{n=0}p_{n}\overline{\alpha}_{s}^{n+1})\,,
\eear
where the first three coefficients are given as \cite{gray,melnikov,chetyrkin}
\bear
p_{0}=0.4244\,,\,\,\,p_{1}=1.3621-0.1055 N_{f}\,,\,\,\,p_{2}=6.1404-0.8597N_{f}+0.0211N_{f}^{2}\,.
\eear

The normalization $\cal N$ is  given by
\bear
{\cal N}=R(1/2)\,,
\eear
where
\bear
R(b)=\tilde m(b)(1-2b)^{1+\nu}\,,
\label{rb}
\eear
and expanding $R(b)$ at the origin using (\ref{pertBorel})
the normalization can be evaluated perturbatively \cite{lee1,lee2}.
While this yields a convergent series for the normalization
 it does not fully exploit the
bilocal expansion (\ref{bilocal}), especially 
the expansion about the renormalon singularity, of which
the coefficients $c_{i}$ are entirely dependent 
on the beta function \cite{bene}, and with four-loop
beta function the first two coefficients $c_{1},c_{2}$ are known \cite{pineda}.
To utilize this expansion we write $R(b)$ in a truncated form:
\bear
R(b)=\left[\sum_{n=0}^{1}\frac{h_{n}}{n!}\left(\frac{b}
{\beta_{0}}\right)^{n}\right](1-2b)^{1+\nu}+{\cal N}[1+\sum_{i=1}^{2} c_{i}(1-2b)^{i}]\,,
\label{truncR}
\eear
and determine $h_{0},h_{1},{\cal N}$ by demanding (\ref{truncR})
 and (\ref{rb}) with (\ref{pertBorel}) give identical 
expansion about the origin to ${\cal O}(b^{2})$. 
To get the numbers in Table \ref{table1} this procedure was
 performed not in the $b-$plane but in the
conformally mapped $z$-plane defined by
\bear
z=\frac{b}{1+b}\,,
\eear
where the main advantage of this mapping is that the first 
UV renormalon at $b=-1$ is mapped away to infinity.
The first IR renormalon is now at $z=1/3$ and 
the bilocal expansion of $R$ in $z$-plane is given by
\bear
R(b(z))=\left[\sum_{n=0}^{1}\frac{h^{z}_{n}}{n!}\left(\frac{z}{\beta_{0}}\right)^{n}\right]
(1-3z)^{1+\nu}+{\cal N}
[1+\sum_{i=1}^{2} c^{z}_{i}(1-3z)^{i}]\,,
\label{truncRz}
\eear
from which ${\cal N}_{\rm pert.}$ in Table \ref{table1} were obtained. Note that 
$c^{z}_{1},c^{z}_{2}$ are given in terms of $c_{1},c_{2}$ by
\bear
c^{z}_{1}=\frac{3}{2} c_{1}\,,\,\,\, c^{z}_{2}=\frac{9}{4} c_{2}-\frac{3}{4} c_{1}\,.
\eear

In summary, we have proposed an ansatz for 
the normalization constant for the first IR renormalon in the pole mass and
presented an argument that, at least,
 (\ref{singamma}) should be an integral part of the normalization constant.


\begin{acknowledgments}
I thank S. Han for communications.
This work was supported by Basic Science
Research Program of the National Research Foundation of Korea
(NRF), funded by  the Ministry of Education, Science and Technology
(2012R1A1A2044543).

\end{acknowledgments}

\bibliographystyle{apsrev4-1}
\bibliography{flavordependence}

\begin{thebibliography}{25}%
\makeatletter
\providecommand \@ifxundefined [1]{%
 \@ifx{#1\undefined}
}%
\providecommand \@ifnum [1]{%
 \ifnum #1\expandafter \@firstoftwo
 \else \expandafter \@secondoftwo
 \fi
}%
\providecommand \@ifx [1]{%
 \ifx #1\expandafter \@firstoftwo
 \else \expandafter \@secondoftwo
 \fi
}%
\providecommand \natexlab [1]{#1}%
\providecommand \enquote  [1]{``#1''}%
\providecommand \bibnamefont  [1]{#1}%
\providecommand \bibfnamefont [1]{#1}%
\providecommand \citenamefont [1]{#1}%
\providecommand \href@noop [0]{\@secondoftwo}%
\providecommand \href [0]{\begingroup \@sanitize@url \@href}%
\providecommand \@href[1]{\@@startlink{#1}\@@href}%
\providecommand \@@href[1]{\endgroup#1\@@endlink}%
\providecommand \@sanitize@url [0]{\catcode `\\12\catcode `\$12\catcode
  `\&12\catcode `\#12\catcode `\^12\catcode `\_12\catcode `\%12\relax}%
\providecommand \@@startlink[1]{}%
\providecommand \@@endlink[0]{}%
\providecommand \url  [0]{\begingroup\@sanitize@url \@url }%
\providecommand \@url [1]{\endgroup\@href {#1}{\urlprefix }}%
\providecommand \urlprefix  [0]{URL }%
\providecommand \Eprint [0]{\href }%
\providecommand \doibase [0]{http://dx.doi.org/}%
\providecommand \selectlanguage [0]{\@gobble}%
\providecommand \bibinfo  [0]{\@secondoftwo}%
\providecommand \bibfield  [0]{\@secondoftwo}%
\providecommand \translation [1]{[#1]}%
\providecommand \BibitemOpen [0]{}%
\providecommand \bibitemStop [0]{}%
\providecommand \bibitemNoStop [0]{.\EOS\space}%
\providecommand \EOS [0]{\spacefactor3000\relax}%
\providecommand \BibitemShut  [1]{\csname bibitem#1\endcsname}%
\let\auto@bib@innerbib\@empty
\bibitem [{\citenamefont {Parisi}(1979)}]{parisi0}%
  \BibitemOpen
  \bibfield  {author} {\bibinfo {author} {\bibfnamefont {G.}~\bibnamefont
  {Parisi}},\ }\href {\doibase 10.1016/0550-3213(79)90298-0} {\bibfield
  {journal} {\bibinfo  {journal} {Nucl.Phys.}\ }\textbf {\bibinfo {volume}
  {B150}},\ \bibinfo {pages} {163} (\bibinfo {year} {1979})}\BibitemShut
  {NoStop}%
\bibitem [{\citenamefont {Parisi}(1978)}]{parisi1}%
  \BibitemOpen
  \bibfield  {author} {\bibinfo {author} {\bibfnamefont {G.}~\bibnamefont
  {Parisi}},\ }\href {\doibase 10.1016/0370-2693(78)90101-6} {\bibfield
  {journal} {\bibinfo  {journal} {Phys.Lett.}\ }\textbf {\bibinfo {volume}
  {B76}},\ \bibinfo {pages} {65} (\bibinfo {year} {1978})}\BibitemShut
  {NoStop}%
\bibitem [{\citenamefont {David}(1984)}]{david1}%
  \BibitemOpen
  \bibfield  {author} {\bibinfo {author} {\bibfnamefont {F.}~\bibnamefont
  {David}},\ }\href {\doibase 10.1016/0550-3213(84)90235-9} {\bibfield
  {journal} {\bibinfo  {journal} {Nucl.Phys.}\ }\textbf {\bibinfo {volume}
  {B234}},\ \bibinfo {pages} {237} (\bibinfo {year} {1984})}\BibitemShut
  {NoStop}%
\bibitem [{\citenamefont {David}(1986)}]{david2}%
  \BibitemOpen
  \bibfield  {author} {\bibinfo {author} {\bibfnamefont {F.}~\bibnamefont
  {David}},\ }\href {\doibase 10.1016/0550-3213(86)90279-8} {\bibfield
  {journal} {\bibinfo  {journal} {Nucl.Phys.}\ }\textbf {\bibinfo {volume}
  {B263}},\ \bibinfo {pages} {637} (\bibinfo {year} {1986})}\BibitemShut
  {NoStop}%
\bibitem [{\citenamefont {Novikov}\ \emph {et~al.}(1985)\citenamefont
  {Novikov}, \citenamefont {Shifman}, \citenamefont {Vainshtein},\ and\
  \citenamefont {Zakharov}}]{nsvz1}%
  \BibitemOpen
  \bibfield  {author} {\bibinfo {author} {\bibfnamefont {V.}~\bibnamefont
  {Novikov}}, \bibinfo {author} {\bibfnamefont {M.~A.}\ \bibnamefont
  {Shifman}}, \bibinfo {author} {\bibfnamefont {A.}~\bibnamefont {Vainshtein}},
  \ and\ \bibinfo {author} {\bibfnamefont {V.~I.}\ \bibnamefont {Zakharov}},\
  }\href {\doibase 10.1016/0550-3213(85)90087-2} {\bibfield  {journal}
  {\bibinfo  {journal} {Nucl.Phys.}\ }\textbf {\bibinfo {volume} {B249}},\
  \bibinfo {pages} {445} (\bibinfo {year} {1985})}\BibitemShut {NoStop}%
\bibitem [{\citenamefont {Novikov}\ \emph {et~al.}(1984)\citenamefont
  {Novikov}, \citenamefont {Shifman}, \citenamefont {Vainshtein},\ and\
  \citenamefont {Zakharov}}]{nsvz2}%
  \BibitemOpen
  \bibfield  {author} {\bibinfo {author} {\bibfnamefont {V.}~\bibnamefont
  {Novikov}}, \bibinfo {author} {\bibfnamefont {M.~A.}\ \bibnamefont
  {Shifman}}, \bibinfo {author} {\bibfnamefont {A.}~\bibnamefont {Vainshtein}},
  \ and\ \bibinfo {author} {\bibfnamefont {V.~I.}\ \bibnamefont {Zakharov}},\
  }\href {\doibase 10.1016/0370-1573(84)90021-8} {\bibfield  {journal}
  {\bibinfo  {journal} {Phys.Rept.}\ }\textbf {\bibinfo {volume} {116}},\
  \bibinfo {pages} {103} (\bibinfo {year} {1984})}\BibitemShut {NoStop}%
\bibitem [{\citenamefont {Mueller}(1985)}]{mueller}%
  \BibitemOpen
  \bibfield  {author} {\bibinfo {author} {\bibfnamefont {A.~H.}\ \bibnamefont
  {Mueller}},\ }\href {\doibase 10.1016/0550-3213(85)90485-7} {\bibfield
  {journal} {\bibinfo  {journal} {Nucl.Phys.}\ }\textbf {\bibinfo {volume}
  {B250}},\ \bibinfo {pages} {327} (\bibinfo {year} {1985})}\BibitemShut
  {NoStop}%
\bibitem [{\citenamefont {Lee}(1997)}]{lee1}%
  \BibitemOpen
  \bibfield  {author} {\bibinfo {author} {\bibfnamefont {T.}~\bibnamefont
  {Lee}},\ }\href {\doibase 10.1103/PhysRevD.56.1091} {\bibfield  {journal}
  {\bibinfo  {journal} {Phys.Rev.}\ }\textbf {\bibinfo {volume} {D56}},\
  \bibinfo {pages} {1091} (\bibinfo {year} {1997})},\ \Eprint
  {http://arxiv.org/abs/hep-th/9611010} {arXiv:hep-th/9611010 [hep-th]}
  \BibitemShut {NoStop}%
\bibitem [{\citenamefont {Lee}(1999)}]{lee2}%
  \BibitemOpen
  \bibfield  {author} {\bibinfo {author} {\bibfnamefont {T.}~\bibnamefont
  {Lee}},\ }\href {\doibase 10.1016/S0370-2693(99)00932-6} {\bibfield
  {journal} {\bibinfo  {journal} {Phys.Lett.}\ }\textbf {\bibinfo {volume}
  {B462}},\ \bibinfo {pages} {1} (\bibinfo {year} {1999})},\ \Eprint
  {http://arxiv.org/abs/hep-ph/9908225} {arXiv:hep-ph/9908225 [hep-ph]}
  \BibitemShut {NoStop}%
\bibitem [{\citenamefont {Pineda}(2001)}]{pineda}%
  \BibitemOpen
  \bibfield  {author} {\bibinfo {author} {\bibfnamefont {A.}~\bibnamefont
  {Pineda}},\ }\href {\doibase 10.1088/1126-6708/2001/06/022} {\bibfield
  {journal} {\bibinfo  {journal} {JHEP}\ }\textbf {\bibinfo {volume} {0106}},\
  \bibinfo {pages} {022} (\bibinfo {year} {2001})},\ \Eprint
  {http://arxiv.org/abs/hep-ph/0105008} {arXiv:hep-ph/0105008 [hep-ph]}
  \BibitemShut {NoStop}%
\bibitem [{\citenamefont {Lee}(2003{\natexlab{a}})}]{lee3}%
  \BibitemOpen
  \bibfield  {author} {\bibinfo {author} {\bibfnamefont {T.}~\bibnamefont
  {Lee}},\ }\href {\doibase 10.1103/PhysRevD.67.014020} {\bibfield  {journal}
  {\bibinfo  {journal} {Phys.Rev.}\ }\textbf {\bibinfo {volume} {D67}},\
  \bibinfo {pages} {014020} (\bibinfo {year} {2003}{\natexlab{a}})},\ \Eprint
  {http://arxiv.org/abs/hep-ph/0210032} {arXiv:hep-ph/0210032 [hep-ph]}
  \BibitemShut {NoStop}%
\bibitem [{\citenamefont {Bauer}\ \emph {et~al.}(2012)\citenamefont {Bauer},
  \citenamefont {Bali},\ and\ \citenamefont {Pineda}}]{bali0}%
  \BibitemOpen
  \bibfield  {author} {\bibinfo {author} {\bibfnamefont {C.}~\bibnamefont
  {Bauer}}, \bibinfo {author} {\bibfnamefont {G.~S.}\ \bibnamefont {Bali}}, \
  and\ \bibinfo {author} {\bibfnamefont {A.}~\bibnamefont {Pineda}},\ }\href
  {\doibase 10.1103/PhysRevLett.108.242002} {\bibfield  {journal} {\bibinfo
  {journal} {Phys.Rev.Lett.}\ }\textbf {\bibinfo {volume} {108}},\ \bibinfo
  {pages} {242002} (\bibinfo {year} {2012})},\ \Eprint
  {http://arxiv.org/abs/1111.3946} {arXiv:1111.3946 [hep-ph]} \BibitemShut
  {NoStop}%
\bibitem [{\citenamefont {Bali}\ \emph {et~al.}(2011)\citenamefont {Bali},
  \citenamefont {Bauer},\ and\ \citenamefont {Pineda}}]{bali1}%
  \BibitemOpen
  \bibfield  {author} {\bibinfo {author} {\bibfnamefont {G.~S.}\ \bibnamefont
  {Bali}}, \bibinfo {author} {\bibfnamefont {C.}~\bibnamefont {Bauer}}, \ and\
  \bibinfo {author} {\bibfnamefont {A.}~\bibnamefont {Pineda}},\ }\href@noop {}
  {\bibfield  {journal} {\bibinfo  {journal} {PoS}\ }\textbf {\bibinfo {volume}
  {LATTICE2011}},\ \bibinfo {pages} {222} (\bibinfo {year} {2011})},\ \Eprint
  {http://arxiv.org/abs/1111.6158} {arXiv:1111.6158 [hep-lat]} \BibitemShut
  {NoStop}%
\bibitem [{\citenamefont {Beneke}(1998)}]{beneke4}%
  \BibitemOpen
  \bibfield  {author} {\bibinfo {author} {\bibfnamefont {M.}~\bibnamefont
  {Beneke}},\ }\href {\doibase 10.1016/S0370-2693(98)00741-2} {\bibfield
  {journal} {\bibinfo  {journal} {Phys.Lett.}\ }\textbf {\bibinfo {volume}
  {B434}},\ \bibinfo {pages} {115} (\bibinfo {year} {1998})},\ \Eprint
  {http://arxiv.org/abs/hep-ph/9804241} {arXiv:hep-ph/9804241 [hep-ph]}
  \BibitemShut {NoStop}%
\bibitem [{\citenamefont {Beneke}\ and\ \citenamefont {Braun}(1994)}]{braun}%
  \BibitemOpen
  \bibfield  {author} {\bibinfo {author} {\bibfnamefont {M.}~\bibnamefont
  {Beneke}}\ and\ \bibinfo {author} {\bibfnamefont {V.~M.}\ \bibnamefont
  {Braun}},\ }\href {\doibase 10.1016/0550-3213(94)90314-X} {\bibfield
  {journal} {\bibinfo  {journal} {Nucl.Phys.}\ }\textbf {\bibinfo {volume}
  {B426}},\ \bibinfo {pages} {301} (\bibinfo {year} {1994})},\ \Eprint
  {http://arxiv.org/abs/hep-ph/9402364} {arXiv:hep-ph/9402364 [hep-ph]}
  \BibitemShut {NoStop}%
\bibitem [{\citenamefont {Bigi}\ \emph {et~al.}(1994)\citenamefont {Bigi},
  \citenamefont {Shifman}, \citenamefont {Uraltsev},\ and\ \citenamefont
  {Vainshtein}}]{bigi}%
  \BibitemOpen
  \bibfield  {author} {\bibinfo {author} {\bibfnamefont {I.~I.}\ \bibnamefont
  {Bigi}}, \bibinfo {author} {\bibfnamefont {M.~A.}\ \bibnamefont {Shifman}},
  \bibinfo {author} {\bibfnamefont {N.}~\bibnamefont {Uraltsev}}, \ and\
  \bibinfo {author} {\bibfnamefont {A.}~\bibnamefont {Vainshtein}},\ }\href
  {\doibase 10.1103/PhysRevD.50.2234} {\bibfield  {journal} {\bibinfo
  {journal} {Phys.Rev.}\ }\textbf {\bibinfo {volume} {D50}},\ \bibinfo {pages}
  {2234} (\bibinfo {year} {1994})},\ \Eprint
  {http://arxiv.org/abs/hep-ph/9402360} {arXiv:hep-ph/9402360 [hep-ph]}
  \BibitemShut {NoStop}%
\bibitem [{\citenamefont {Beneke}(1999)}]{beneke1}%
  \BibitemOpen
  \bibfield  {author} {\bibinfo {author} {\bibfnamefont {M.}~\bibnamefont
  {Beneke}},\ }\href {\doibase 10.1016/S0370-1573(98)00130-6} {\bibfield
  {journal} {\bibinfo  {journal} {Phys.Rept.}\ }\textbf {\bibinfo {volume}
  {317}},\ \bibinfo {pages} {1} (\bibinfo {year} {1999})},\ \Eprint
  {http://arxiv.org/abs/hep-ph/9807443} {arXiv:hep-ph/9807443 [hep-ph]}
  \BibitemShut {NoStop}%
\bibitem [{\citenamefont {Beneke}(1995)}]{bene}%
  \BibitemOpen
  \bibfield  {author} {\bibinfo {author} {\bibfnamefont {M.}~\bibnamefont
  {Beneke}},\ }\href {\doibase 10.1016/0370-2693(94)01505-7} {\bibfield
  {journal} {\bibinfo  {journal} {Phys.Lett.}\ }\textbf {\bibinfo {volume}
  {B344}},\ \bibinfo {pages} {341} (\bibinfo {year} {1995})},\ \Eprint
  {http://arxiv.org/abs/hep-ph/9408380} {arXiv:hep-ph/9408380 [hep-ph]}
  \BibitemShut {NoStop}%
\bibitem [{\citenamefont {Lee}(2012)}]{lee-largeorder}%
  \BibitemOpen
  \bibfield  {author} {\bibinfo {author} {\bibfnamefont {T.}~\bibnamefont
  {Lee}},\ }\href {\doibase 10.1016/j.physletb.2012.04.017} {\bibfield
  {journal} {\bibinfo  {journal} {Phys.Lett.}\ }\textbf {\bibinfo {volume}
  {B711}},\ \bibinfo {pages} {360} (\bibinfo {year} {2012})},\ \Eprint
  {http://arxiv.org/abs/1112.4433} {arXiv:1112.4433 [hep-ph]} \BibitemShut
  {NoStop}%
\bibitem [{\citenamefont {Beneke}(1993)}]{beneke3}%
  \BibitemOpen
  \bibfield  {author} {\bibinfo {author} {\bibfnamefont {M.}~\bibnamefont
  {Beneke}},\ }\href {\doibase 10.1016/0550-3213(93)90554-3} {\bibfield
  {journal} {\bibinfo  {journal} {Nucl.Phys.}\ }\textbf {\bibinfo {volume}
  {B405}},\ \bibinfo {pages} {424} (\bibinfo {year} {1993})}\BibitemShut
  {NoStop}%
\bibitem [{\citenamefont {Broadhurst}(1993)}]{broadhurst}%
  \BibitemOpen
  \bibfield  {author} {\bibinfo {author} {\bibfnamefont {D.~J.}\ \bibnamefont
  {Broadhurst}},\ }\href {\doibase 10.1007/BF01560355} {\bibfield  {journal}
  {\bibinfo  {journal} {Z.Phys.}\ }\textbf {\bibinfo {volume} {C58}},\ \bibinfo
  {pages} {339} (\bibinfo {year} {1993})}\BibitemShut {NoStop}%
\bibitem [{\citenamefont {Lee}(2003{\natexlab{b}})}]{lee4}%
  \BibitemOpen
  \bibfield  {author} {\bibinfo {author} {\bibfnamefont {T.}~\bibnamefont
  {Lee}},\ }\href@noop {} {\bibfield  {journal} {\bibinfo  {journal} {JHEP}\
  }\textbf {\bibinfo {volume} {0310}},\ \bibinfo {pages} {044} (\bibinfo {year}
  {2003}{\natexlab{b}})},\ \Eprint {http://arxiv.org/abs/hep-ph/0304185}
  {arXiv:hep-ph/0304185 [hep-ph]} \BibitemShut {NoStop}%
\bibitem [{\citenamefont {Gray}\ \emph {et~al.}(1990)\citenamefont {Gray},
  \citenamefont {Broadhurst}, \citenamefont {Grafe},\ and\ \citenamefont
  {Schilcher}}]{gray}%
  \BibitemOpen
  \bibfield  {author} {\bibinfo {author} {\bibfnamefont {N.}~\bibnamefont
  {Gray}}, \bibinfo {author} {\bibfnamefont {D.~J.}\ \bibnamefont
  {Broadhurst}}, \bibinfo {author} {\bibfnamefont {W.}~\bibnamefont {Grafe}}, \
  and\ \bibinfo {author} {\bibfnamefont {K.}~\bibnamefont {Schilcher}},\ }\href
  {\doibase 10.1007/BF01614703} {\bibfield  {journal} {\bibinfo  {journal}
  {Z.Phys.}\ }\textbf {\bibinfo {volume} {C48}},\ \bibinfo {pages} {673}
  (\bibinfo {year} {1990})}\BibitemShut {NoStop}%
\bibitem [{\citenamefont {Melnikov}\ and\ \citenamefont
  {Ritbergen}(2000)}]{melnikov}%
  \BibitemOpen
  \bibfield  {author} {\bibinfo {author} {\bibfnamefont {K.}~\bibnamefont
  {Melnikov}}\ and\ \bibinfo {author} {\bibfnamefont {T.~v.}\ \bibnamefont
  {Ritbergen}},\ }\href {\doibase 10.1016/S0370-2693(00)00507-4} {\bibfield
  {journal} {\bibinfo  {journal} {Phys.Lett.}\ }\textbf {\bibinfo {volume}
  {B482}},\ \bibinfo {pages} {99} (\bibinfo {year} {2000})},\ \Eprint
  {http://arxiv.org/abs/hep-ph/9912391} {arXiv:hep-ph/9912391 [hep-ph]}
  \BibitemShut {NoStop}%
\bibitem [{\citenamefont {Chetyrkin}\ and\ \citenamefont
  {Steinhauser}(2000)}]{chetyrkin}%
  \BibitemOpen
  \bibfield  {author} {\bibinfo {author} {\bibfnamefont {K.}~\bibnamefont
  {Chetyrkin}}\ and\ \bibinfo {author} {\bibfnamefont {M.}~\bibnamefont
  {Steinhauser}},\ }\href {\doibase 10.1016/S0550-3213(99)00784-1} {\bibfield
  {journal} {\bibinfo  {journal} {Nucl.Phys.}\ }\textbf {\bibinfo {volume}
  {B573}},\ \bibinfo {pages} {617} (\bibinfo {year} {2000})},\ \Eprint
  {http://arxiv.org/abs/hep-ph/9911434} {arXiv:hep-ph/9911434 [hep-ph]}
  \BibitemShut {NoStop}%
\end{thebibliography}%

\end{document}